\documentclass[journal=cmatex,manuscript=article]{achemso}

\usepackage{chemformula} 
\usepackage[T1]{fontenc} 
\usepackage{lineno}
\usepackage{hyperref}
\usepackage{makecell}

\makeatletter
\newcommand{\customlabel}[2]{%
   \protected@write \@auxout {}{\string \newlabel {#1}{{#2}{\thepage}{#2}{#1}{}} }%
   \hypertarget{#1}{#2}
}
\makeatother

\usepackage{xcolor,soul}
\newcommand\nothing{}
\renewcommand{\hl}{\nothing} 

\author{Kazuki Morita}
\affiliation{Department of Chemistry, University of Pennsylvania, Philadelphia, Pennsylvania 19104-6323, USA}
\author{Andrew M. Rappe}
\affiliation{Department of Chemistry, University of Pennsylvania, Philadelphia, Pennsylvania 19104-6323, USA}
\email{rappe@sas.upenn.edu}

\title{
        Inverted band gap trend through octahedral ordering in \ch{Cs2Au2X6} (X=Cl, Br, I)
}

\begin{document}

%
%

\begin{abstract}
Double perovskites \ch{Cs2Au2X6} (X=Cl, Br, I) are prototypical materials that exhibit charge disproportionation of gold into 1+ and 3+ states. It is known that the disproportionation is resolved under high pressures, and this has stimulated many studies into the pressurization of these materials. At present, the phase changes in these materials are still strongly contested. Here, we use density functional theory to study the pressure-dependent behavior of \ch{Cs2Au2X6}. We find that a tetragonal--cubic transition occurs directly from the ground state $I4/mmm$ structure. Even so, we also found an intermediate tetragonal $P4/mmm$ structure to be very close in energy, suggesting it to be observable. We also find several other competing metastable phases, which explains some of the controversies in the literature. Focusing on one of the metastable phases, we suggest that \ch{Cs2Au2X6} can be prepared in a $P4_2/mnm$ structure, analogous to that of \ch{KCuF3}. The band gap in the $P4_2/mnm$ structure widened as atomic number of the halide was increased, which is the inverse trend compared to the ground state structure. We explain this by the different octahedral distortion ordering in the two structural phases. Furthermore, we show that the conduction band in $P4_2/mnm$ is three dimensionally connected, which is favorable for opto-electronic applications. We submit that this work demonstrates that octahedral distortion ordering is a promising avenue for developing new double perovskites and suggests it to be particular effective in tuning the electronic structure properties.
\end{abstract}


\section{Introduction}

Halide perovskites have attracted considerable attention in recent years as a promising class of materials for optoelectronic applications.\cite{Brenner2016,Sendner2016,Nayak2019,Kaiser2022}
Within this class, \ch{Cs2Au2X6} (X=Cl, Br, I) is particularly interesting, due to it crystallizing in a double perovskite structure despite having a ternary composition.
This change is driven by the disproportionation of gold into oxidation states 1+ and 3+.\cite{Denner1979,Kojima1994,Ushakov2011}
Study of charge disproportionation has been an active field and has become one of the major design principles to access electronic properties that are otherwise inaccessible.\cite{Mizokawa1991,Benam2021,WuJ2021,Zunger2021,Bennett2022}
The disproportionation accompanies octahedral distortion, where \ch{Au^{1+}X6} and \ch{Au^{3+}X6} exhibit compressed and elongated octahedra along the $c$ axis, respectively.
\ch{Cs2Au2X6} is insulating in the ground state and exhibits a distinctive fully isolated conduction band edge, which is usually assigned to the unoccupied gold $5d_{x^2-y^2}$ orbital in \ch{Au^{3+}}.\cite{LiuX1999}
The transition from the valence band to the conduction band is optically allowed, which reduces the absorption edge substantially and has stimulated studies to use these materials as photovoltaics.\cite{Debbichi2018,Giorgi2018,HuangY2020,Kangsabanik2020,Bhawna2021,ZhangP2021}

From very early studies, dissolution of the gold charge ordering under pressure was known and was experimentally investigated.\cite{Denner1979,Kojima1990}
The initial studies reported that these materials undergo a tetragonal--tetragonal insulator--metal transition ($I4/mmm$ (No. 139) to $P4/mmm$ (No. 123)) and a subsequent tetragonal--cubic metal--metal transition ($P4/mmm$ (No. 123) to $Pm\overline{3}m$ (No. 221)).\cite{Kojima1990,Kojima1994}
The two transitions occur at 4.5 and 6.5 GPa for \ch{Cs2Au2I6}.\cite{Kojima1990}
Since then, \ch{Cs2Au2X6} has served as a model system to study the coupling between the lattice and orbital degrees of freedom.
However, more recently, a series of different phase transitions was reported that challenged the original consensus.\cite{Kusmartseva2010,WangS2013,WangS2014}
For example, Attfield and co-workers have reported that the tetragonal--tetragonal transition ($I4/mmm$ to $P4/mmm$) is an insulator--insulator transition and that amorphization occurs before the insulator--metal transition ($P4/mmm$ to $Pm\overline{3}m$).\cite{Kusmartseva2010}
More recently, Wang and co-workers have suggested that a tetragonal--orthorhombic transition occurs instead of the tetragonal--tetragonal transition, and the material subsequently becomes amorphous before becoming cubic.\cite{WangS2013}
They also suggested that metallic behavior is only observed near the tetragonal--orthorhombic transition, and that the band gap reopens at higher pressure.\cite{WangS2014}
The newly proposed $Ibmm$ (alternative axes of $Imma$ No. 74) orthorhombic phase also has octahedral tilting, which was not considered previously.
On the other hand, computational work yielded an inconclusive result and has suggested that charge reordering in \ch{Cs2Au2Cl6} occurs at the first phase transition but could not identify the high-pressure phase due to subtle differences between the tetragonal $P4/mmm$ and cubic $Pm\overline{3}m$ structures.\cite{Winkler2001}
These works show that phase transition in \ch{Cs2Au2X6} is still an open question, and the conflicting claims could potentially imply the existence of multiple metastable phases.

Optoelectronic properties of most halide perovskites are determined by the B- and X-site composition, and the \ch{BX6} octahedral distortion strongly affects the band gap.\cite{ZhaoXG2018,Marchenko2021,Wolf2021,Morita2022_CM}
\hl{
        (pseudo-)Jahn-Teller distortions, which are often present in {\ch{BX6}} octahedra with transition metal B-sites, can cause elongation or compression of the octahedra.{\cite{Bersuker1975,Bersuker2001}}
}
Since the X-site is shared by a pair of neighboring \ch{BX6}, the octahedral distortion cannot happen independently and the collective orbital ordering can lead to many different crystallographic structures, with a wide range of chemical properties.\cite{Wolf2021,Goodenough2004}
For example, Wei and co-workers have comprehensively studied A- and B-site alloying using first-principles calculations.\cite{HuangY2020}
They report band bending and suggest that alloying in halide perovskites does not simply provide an average of ternary perovskite properties, but instead realizes properties beyond what is seen in ternary perovskites.
Not only the band gap size but the location of the band edges within the Brillouin zone have been reported to change with alloying.\cite{ZhaoXG2018}
In \ch{Cs2Tl2Cl6}, octahedral distortion accompanied by charge disproportionation of Tl to \ch{Tl^{1+}} and \ch{Tl^{3+}} was suggested to realize superconductivity.\cite{Retuerto2013}
More recently, \ch{Cs4Au3Cl12}, a gold-defective \ch{CsAuCl3}, was synthesized, where the absence of the neighboring Au allows for an Au--Cl bond elongation that can host a novel charge state of 2+ in gold.\cite{Lindquist2023}
In \ch{KCuF3}, multiple octahedral distortion patterns are very close in energy, causing multiple metastable structures to be observed experimentally.\cite{Tsukuda1972}
The Cu orbital ordering tunes the superexchange coupling and can host numerous different spin orientations.\cite{Towler1995,Curely2021,Moreira1999,Paolasini2002,Scheie2021}
The above works highlight the importance of octahedral distortion and their ordering and indicate that their tuning is a promising approach to altering the electronic properties.

In this work, we used density functional theory (DFT) to investigate the high-pressure phases of \ch{Cs2Au2X6} (X=Cl, Br, I).
Our results show that direct transition to the cubic $Pm\overline{3}m$ structure is possible and is accompanied by metal--insulator transition.
The experimentally observed tetragonal $P4/mmm$ structure is very close in energy, suggesting that it can be observed.
The calculations do not support the presence of either an intermediate orthorhombic phase or octahedral tilting.
The nature of the phase transition also differs, depending on the choice of halide, suggesting the strong influence of the effective ion size.
During these calculations, we observed multiple competing phases that may have been observed in some experiments.
Investigating one of these phases more deeply, we found that the octahedral distortion pattern that is analogous to the \ch{KCuF3} structure can be metastable at ambient pressure.
Despite having the same constituent octahedral building blocks to the ground state $I4/mmm$ structure, the \ch{KCuF3} analog $P4_2/mnm$ structure exhibits different chemical nature of the band-edge states and shows favorable properties for optoelectronic applications.
This work resolves some of the controversies in the literature and suggests octahedral distortion ordering as a promising avenue to tune band structures in double perovskites.

\section{Methodology}
Pseudopotential plane-wave density functional theory calculations were performed using the Quantum Espresso code.\cite{Giannozzi2009,Giannozzi2017}
The PBE exchange-correlation functional was used throughout this work.\cite{Perdew1996}
In addition, calculations based on hybrid exchange-correlation functional HSE06 on HSE06-relaxed structures were also performed as a comparison (Figure~\ref{figSI:ac_hse06}).\cite{Heyd2003}
However, we found that it has a tendency to excessively reduce the density of states near the Fermi-level in the metallic phases, as discussed in the literature.\cite{GaoW2016}
A 40-atom supercell was used for pressurization and band structure calculation to allow comparison between different crystal space groups.
Phonon modes were calculated with a finite displacement method in a 320-atom supercell using the {\it phonopy} package.\cite{Togo2023_JPCM,Togo2023_JPSJ}
A Monkhorst-Pack sampling was adopted, with sampling density of at most 0.15~\AA$^{-1}$\ between the neighboring k-points and a plane-wave cutoff of at least 80 Ry.
The large cutoff energy was selected in order to ensure stringent convergence, even though the pseudopotentials can ordinarily be used with a small cutoff.
Norm-conserving pseudopotentials were generated using the OPIUM package and the RRKJ method.\cite{Rappe1990}
Since the inter-atomic distances are small when the crystal is under pressure, small pseudopotential cutoff radii of 1.85, 2.07, 2.0, 1.8, and 2.0 Bohr was used for Cs, Au, Cl, Br, and I, respectively.

To calculate the pressure dependence of the structural and electronic properties, full structural relaxation was first performed using the BFGS algorithm to optimize all the atomic positions and all unit cell parameters.
To model pressure dependence, the volumes of these structures were scaled, and subsequent constant-volume relaxations were performed, allowing all axes and atoms to relax while constraining the volume.
Even though we started with the equilibrium ground-state structure, we found many cases where the structure relaxation at reduced volume can become trapped in a local minimum, with other lower-energy structural minima possible.
To remove bias and find all the relevant energy minima, we classified the relaxed structures according to their structural phase and performed interpolation of the structures, such that all the phases found at any volume are relaxed and identified for all volumes.
We removed clearly unphysical or unstable structures, and the resulting structures are presented in Figure~\ref{figSI:struct_prerelax}.
This interpolation allowed us to obtain many starting structures for any given volume, for which constant-volume relaxations were performed.
We report the energies of the final relaxed structures in all identified phases.

\section{Results and Dicussions}
\subsection{Ground state properties}


\begin{table}[h!]
        \caption{Bond lengths, volumes, and band gaps in \ch{Cs2Au2X6} (X = Cl, Br, I) ground state $I4/mmm$ structure.}
  \label{table:bonding_vol_bg}
        \begin{tabular}{ |c|c|c|c|c|c|c|}
 \hline
        X &
        \makecell{nearest \\ \ch{Au^{1+}}-X (\AA)} & 
        \makecell{second nearest \\ \ch{Au^{1+}}-X (\AA)} & 
        \makecell{nearest \\ \ch{Au^{3+}}-X (\AA)} & 
        \makecell{second nearest \\ \ch{Au^{3+}}-X (\AA)} &
                \makecell{volume \\(\AA$^3$)} &
                \makecell{HSE06 \\ band gap (eV)} \\
 \hline
                Cl & 2.34 & 3.01 & 2.38 & 3.32 & 164.82 & 1.85 \\
 \hline
                Br & 2.49 & 3.08 & 2.55 & 3.43 & 188.03 & 1.64 \\
 \hline
                I  & 2.66 & 3.26 & 2.74 & 3.62 & 226.47 & 1.57 \\
 \hline
\end{tabular}
\end{table}

\begin{figure}[h!tb]
\includegraphics[width=140mm]{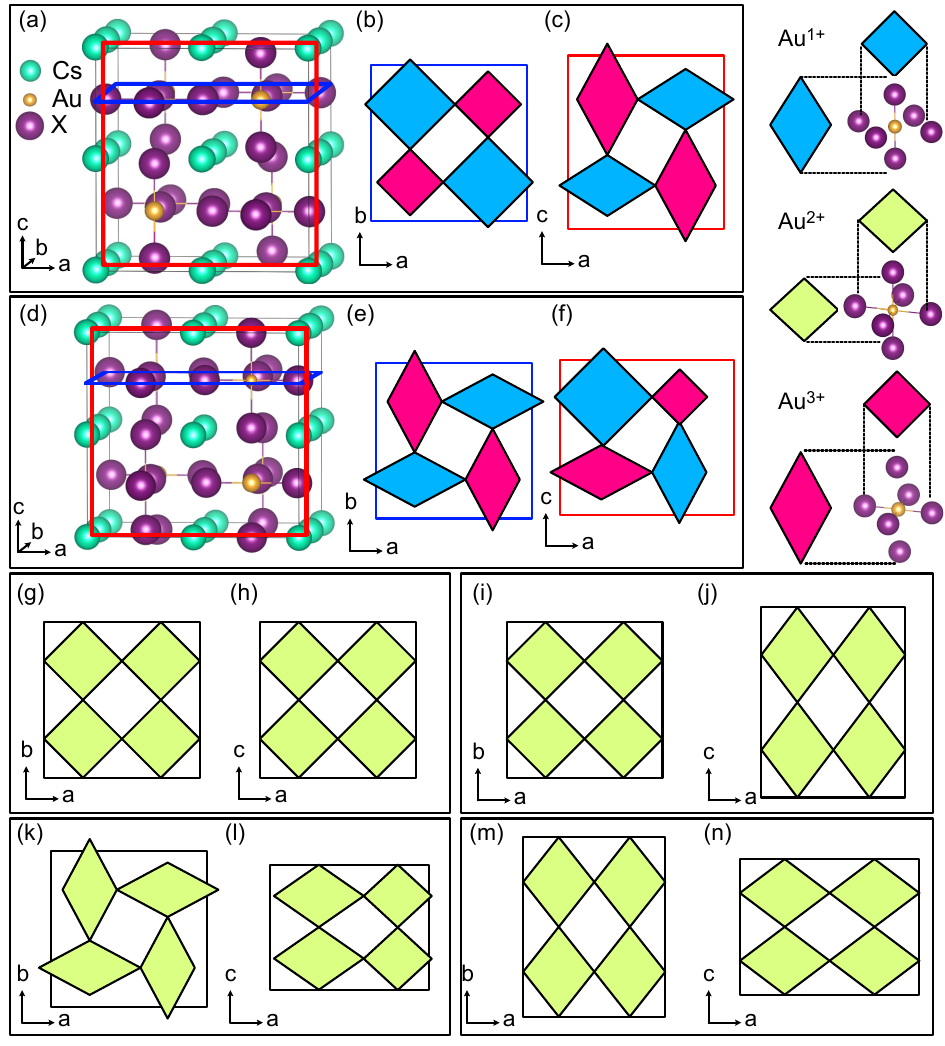}
\caption{\label{fig:struct}
   Phases considered in this study.
        The blue, lime, and magenta quadrilaterals represents \ch{Au^{1+}}, \ch{Au^{2+}}, and \ch{Au^{3+}} sites, respectively.
        The octahedral orientation is represented by square (rhombus) if viewed parallel (orthogonal) to the four-fold rotation axis (schematically shown on the right of (a)).
        Note that \ch{Au^{2+}} only appears in metallic phases and thus is only used as a label rather than to specify an oxidation state.
        Also, \ch{Au^{2+}} can be a perfect regular octahedron, which is represented by a square for all directions.
        (a) and (d) show the atomic structure, where the green, yellow, and purple circles represent cesium, gold, and halide atoms, respectively.
        (b), (c), (e), (f), and (g)-(n) show the schematic orientation of \ch{AuX6} octahedra.
        (a)-(c) Ground-state tetragonal $I4/mmm$ (No. 139),
        (d)-(f) \ch{KCuF3} analog structure $P4_2/mnm$ (No. 136),
        (g)-(h) ideal perovskite cubic $Pm\overline{3}m$ (No. 221),
        (i)-(j) tetragonal $P4/mmm$ (No. 123),
        (k)-(l) tetragonal $P4/mbm$ (No. 127), and
        (m)-(n) orthorhombic $Pmma$ (No. 51).
  }
\end{figure}

\begin{figure}[h!tb]
\includegraphics[width=100mm]{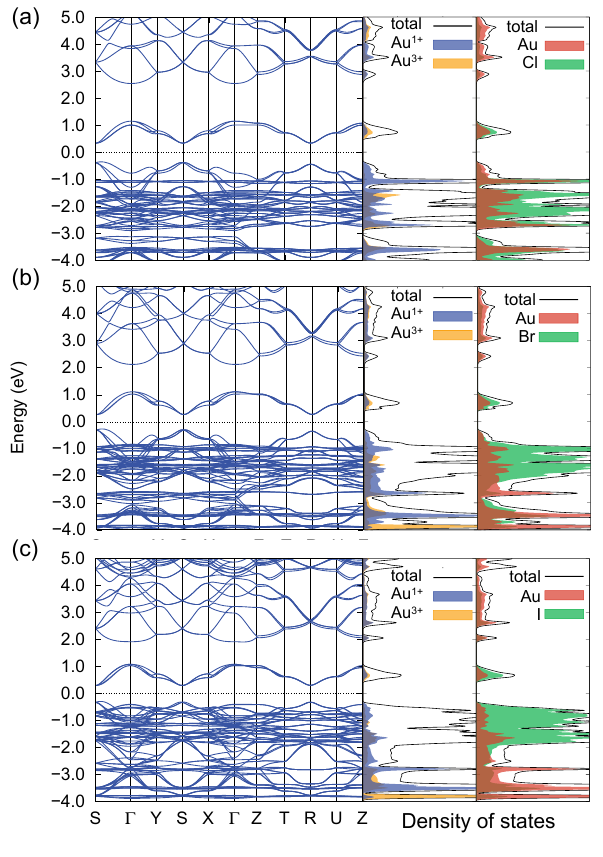}
\caption{\label{fig:elec}
        Band structure and density of states of (a) \ch{Cs2Au2Cl6}, (b) \ch{Cs2Au2Br6}, and (c) \ch{Cs2Au2I6}.
  The zero of energy is taken to be the Fermi energy.
  }
\end{figure}

The ground-state structure of \ch{Cs2Au2X6} (X=Cl, Br, I) was obtained by fully relaxing a 40-atom unit cell (Figures~\ref{fig:struct}(a)-(c)).
Although the reported phases of \ch{Cs2Au2X6} can mostly be treated with 10-atom cells, this choice was made in order to consider various octahedral tilting and spin patterns.
We confirmed that the ground state is $I4/mmm$ with gold disproportionation, in agreement with previous studies.\cite{Kojima1994,Debbichi2018}
The bond lengths are summarized in Table~\ref{table:bonding_vol_bg}.
The elongation in \ch{Au^{1+}} is about 25\%, whereas elongation in \ch{Au^{3+}} is about 35\%.
The lengths are in agreement with previous computational work,\cite{Winkler2001} but overestimate the experimental bond length.\cite{Kusmartseva2010,Riggs2012,Denner1979}
The calculated ground-state volumes are shown in Table~\ref{table:bonding_vol_bg}.
To avoid confusion, energies and volumes are reported per primitive formula unit of the cubic perovskite (\ch{CsAuX3}) throughout the rest of the work.
The electronic band structure (Figure~\ref{fig:elec}) exhibits a distinctive set of bands that is isolated from the rest of the conduction bands throughout the Brillouin zone.
Both PBE and HSE06 functionals consistently showed that the band gap narrows as atomic number of the halide is increased (Table~\ref{table:bonding_vol_bg}).
All systems did not exhibit local or global spin polarization.
The $d^8$ configuration in an octahedral coordination environment like \ch{Au^{3+}} can exhibit high-spin configuration, but as is often seen in 5$d$ orbital systems, the low-spin state is preferred.
The singlet ground state suggests that the energy gain of taking a singlet configuration and lowering the energy level of $d_{z^2}$ and raising $d_{x^2-y^2}$ by distortion of the octahedra is more favorable than taking the triplet configuration without any octahedral distortion.
The valence band mostly consists of halide p-orbitals, and the main conduction band is largely composed of gold 6s-orbitals (Figure~\ref{fig:elec}).
\ch{Cs2Au2I6} has more halide contribution in the valence band compared to \ch{Cs2Au2Cl6} and \ch{Cs2Au2Br6}.
The isolated band is a hybridization of gold and halide states.
Remarkably, the isolated band also shows sizable contributions from \ch{Au^{1+}}, which suggests that the gold is not in the idealized $5d^8$ or $5d^{10}$ electronic configuration, and that this simplified understanding may have limited validity.

\subsection{Pressure dependence}
\begin{figure}[h!tb]
\includegraphics[width=170mm]{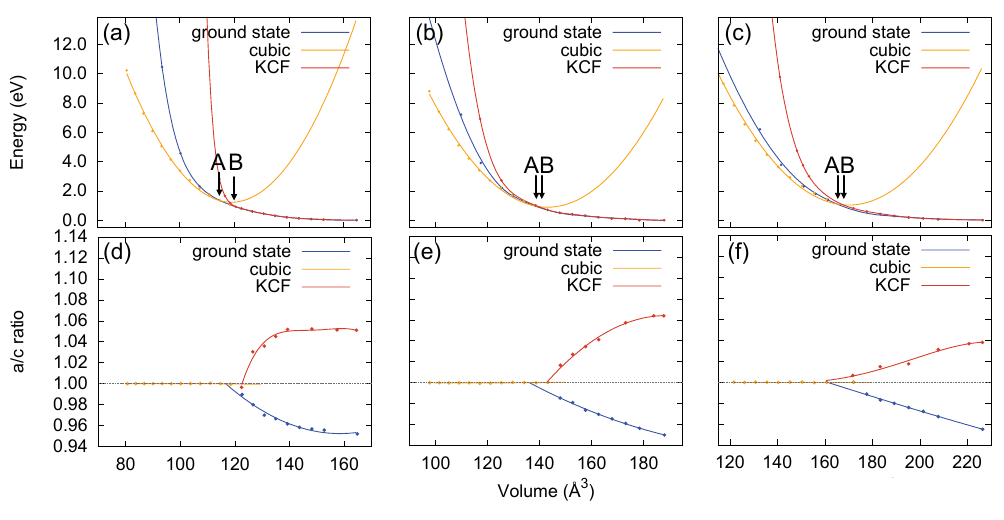}
\caption{\label{fig:ac_ene}
 Energy change with volume for (a) \ch{Cs2Au2Cl6}, (b) \ch{Cs2Au2Br6}, and (c) \ch{Cs2Au2I6}.
        Ground state, cubic, and KCF correspond to the $I4/mmm$, $Pm\overline{3}m$, and $P4_2/mnm$ structures, respectively.
        The arrows ``A'' and ``B'' indicate the transition point for the $I4/mmm$ and the $P4_2/mnm$ structures, respectively.
 $a/c$ ratio change with volume for (d) \ch{Cs2Au2Cl6}, (e) \ch{Cs2Au2Br6}, and (f) \ch{Cs2Au2I6}.
        The spline lines are provided as the guide for a eye.
  }
\end{figure}

\begin{figure}[h!tb]
\includegraphics[width=70mm]{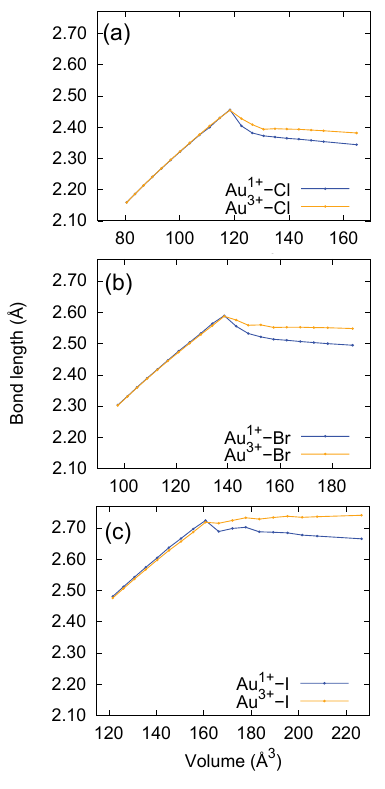}
\caption{\label{fig:fig_bonds}
        Nearest neighbor bond length changes when the ground-state tetragonal structure is compressed for (a) \ch{Cs2Au2Cl6}, (b) \ch{Cs2Au2Br6}, and (c) \ch{Cs2Au2I6}.
        Results on next-nearest-neighbor distance are shown in Figure~\ref{figSI:v_bond}.
  }
\end{figure}

Figures~\ref{fig:ac_ene}(a)-(c) show the energy change with volume change for \ch{Cs2Au2Cl6}, \ch{Cs2Au2Br6}, and \ch{Cs2Au2I6}, respectively.
The zero of energy is taken as the energy of the ground state $I4/mmm$ structure under no pressure.
Since the $I4/mmm$ structure relaxes spontaneously to the cubic $Pm\overline{3}m$ phase, the points of $I4/mmm$ structure above the energy of the $Pm\overline{3}m$ structure were fixed from the extrapolated value and were not relaxed.
This allowed us to define the $I4/mmm$ energy volume curve beyond the transition point and obtain a smooth curve near the transition point.
In \ch{Cs2Au2X6} (X=Cl, Br and I), the $I4/mmm$ tetragonal to the $Pm\overline{3}m$ cubic structure phase transition occurred at volumes of 115, 138, and 166 {~\AA$^3$} (16.6, 9.7 and, 8.9 GPa (Figure~\ref{figSI:e_p})), respectively.
The energy curve of the ground-state $I4/mmm$ structure was highly non-linear and had a very small gradient.
This reflects the anharmonicity of bonding in halide perovskites,\cite{Tadano2022} and weak bonding in the $c$-direction.
In contrast, the cubic $Pm\overline{3}m$ structure was almost parabolic with a large gradient, suggesting that this phase has higher elastic moduli around the equilibrium volume.

To further understand the transitions from tetragonal to cubic, we analyzed the change in $a/c$ axis ratio with pressure (Figure~\ref{fig:ac_ene}).
In all of the compounds, the $I4/mmm$ tetragonal structure gradually increases its $a/c$ ratio and transforms into the $Pm\overline{3}m$ cubic phase with the gradient of the $a/c$ curve changing discontinuously.
In several instances, when we relax from the ground state $I4/mmm$ structure, the BFGS geometry optimization algorithm often relaxed into the metastable $P4/mmm$ structure (Figures~\ref{fig:struct} (i) and (j)).
For \ch{Cs2Au2Cl6}, the energy difference between the $P4/mmm$ and $Pm\overline{3}m$ phases was only 13.0 and 7.5 meV/f.u. for volumes of 83.65 and 110.97{~\AA$^3$}, respectively, so if the experimental compression was performed quasi-adiabatically, it is likely that the system goes into the $P4/mmm$ phase before going to the cubic phase.
Additionally, our calculation is under ideal condition, without asymmetric strain or defects, so if such an external factor induces asymmetry between the $a$ and $c$ axes, $P4/mmm$ is likely to be favored.
The tetragonal--cubic phase transition has been attributed to a loss of the gold charge ordering and insulator--metal transition.\cite{Kojima1994}
We observe that this happens during the tetragonal--cubic transition (arrow ``A'' in Figure~\ref{fig:ac_ene}).

Interestingly, the curvature of the $a/c$ ratio plot versus pressure within the tetragonal $I4/mmm$ structure (Figures~\ref{fig:ac_ene}(d)-(f)) differed significantly between the halides.
Notably, \ch{Cs2Au2Cl6} exhibited a highly non-linear curve, whereas \ch{Cs2Au2Br6} and \ch{Cs2Au2I6} exhibited nearly linear curves.
The difference can be explained by examining the changes in bond length (Figure~\ref{fig:fig_bonds}).
Because the $d^{10}$ configuration in \ch{Au^{1+}} is spherical and unlikely to drive any distortion, we suggest that the decrease of the $a/c$ ratio in the ground-state tetragonal phase in \ch{Cs2Au2X6} is largely driven by the distortion of the \ch{Au^{3+}} octahedra.
This argument has been suggested previously,\cite{Winkler2001} but was not investigated extensively.
Counter-intuitively, the \ch{Au^{3+}-Cl} bond length increases with increasing pressure in \ch{Cs2Au2Cl6}.
This is strange, considering that the total volume is being reduced suggesting that other bonds are absorbing the volume change.
Analyzing the Cs--Cl bonding (Figure~\ref{figSI:v_bond}(a)), the volume change is largely absorbed by shrinking of the cages around the cesium.
This is possible because the \ch{AuCl6} octahedra are small and in the cases of \ch{AuBr6} or \ch{AuI6}, there is less room to stretch the bonds.
The Cs--X bond length and the longer Au--X bonds in all the halides decreased throughout the compression, which also supports our suggestion.
The trend in the $a/c$ ratio seems to support our hypothesis that \ch{Au^{3+}} is the dominant driver of the tetragonal crystal structure.

The \hl{(pseudo-)Jahn-Teller} distortions led to a series of metastable structures that exhibited different octahedral orientations, as shown in Figure~\ref{fig:struct}.
In addition to the $P4/mmm$ phase discussed above, we also found the orthorhombic $Pmma$ structure to be metastable (Figures~\ref{fig:struct}(m) and (n)).
$Pmma$ phase was a only violation to the octahedral building block picture since the elongation in all three direction differed, thus taking the orthorhombic crystal structure.
Throughout the compression, we did not observe any octahedral tilting, making it distinct from the reported orthorhombic $Ibmm$.\cite{WangS2014}
We also did not observe any signature of amorphization.
This suggests either that amorphization requires defects to form or that a larger calculation cell would be required to reproduce the breaking of crystal periodicity.

Some of the competing phases, e.g. $P4/mmm$, lacked charge disproportionation (Figure~\ref{fig:struct}).
In a simplistic model, these \ch{Au^{2+}} have $d^9$ configuration, and thus should have an unpaired spin.
However, in all of the structures, we did not find any unpaired spin density.
Since DFT is known to be less accurate for describing spins in transition metals, we calculated the same structures using the PBE+$U$ (U=3.4 eV) functional.
The Hubbard $U$ value was adopted from a recent fit using cRPA.\cite{Tesch2022}
The PBE+$U$ approach gave the same result as the standard PBE, and the system was non-magnetic in all cases.
This suggests that gold has a strong preference for remaining in the non-magnetic state.
Compared to copper in \ch{KCuF3}, where it takes oxidation state of 2+, gold has a smaller energy difference between the 6s and 5d orbitals.
This is supported by the fact that the dispersion of the band edges in Figure~\ref{fig:elec_KCF} is larger than the localized d-states near -1.0 eV.
\hl{
        The s-orbital nature of the band-edge states also explains the observed octahedral distortion.
        Under compression, the (pseudo-)Jahn-Teller distortion creates shallow local minima in the potential energy surface which results in multiple metastable structures, such as tetragonal $P4/mmm$ structure with $a/c$ ratio near 1.0.
}
This is likely to be one of the reasons behind the previously noted inconsistency in experimentally assigning space groups.\cite{Kusmartseva2010,WangS2013,WangS2014}
The absence of magnetism also suggests that the highest and lowest energy levels in gold have a sizable contribution from 6s-orbitals.
Compared to 3d-orbitals in copper, the gold orbitals have a much smaller $s$-$d$ energy splitting, and the 6s orbital spatially extends to the neighboring halides.

\subsection{\ch{KCuF3} analog structure}

\begin{figure}[htb]
\includegraphics[width=80mm]{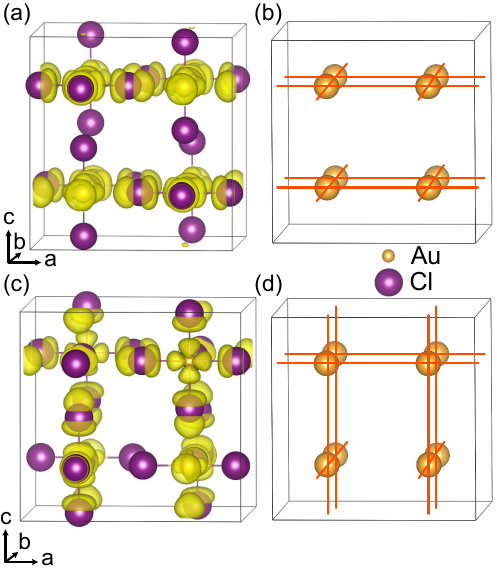}
\caption{\label{fig:band_edges}
        Electronic density of conduction band edge for \ch{Cs2Au2Cl6} in (a) the ground state $I4/mmm$ and in (c) the \ch{KCuF3} analog $P4_2/mnm$ structure.
        The conduction band edge electron networks for (a) and (c) are drawn with orange lines in (b) and (d), respectively.
        Note that Cs is not plotted for visibility.
}
\end{figure}

\begin{figure}[h!tb]
\includegraphics[width=100mm]{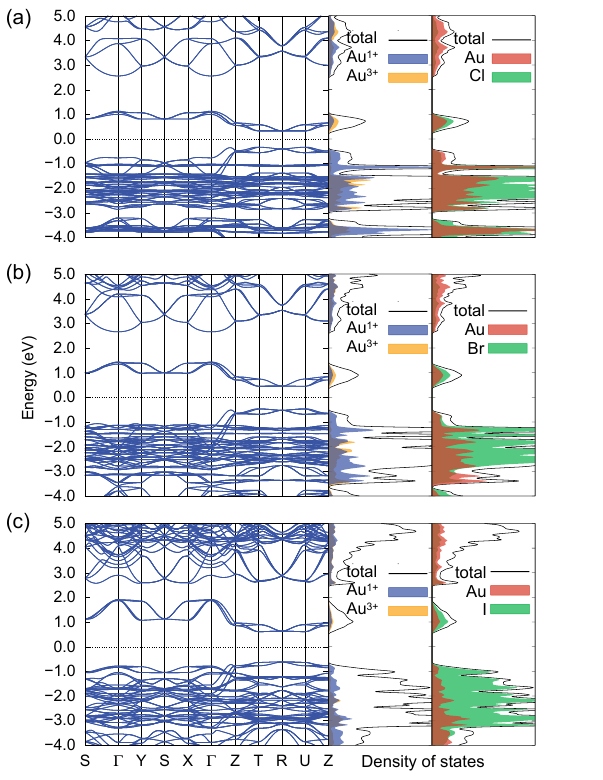}
\caption{\label{fig:elec_KCF}
        Band structure and density of states (a) \ch{Cs2Au2Cl6}, (b) \ch{Cs2Au2Br6}, and (c) \ch{Cs2Au2I6} in \ch{KCuF3} analog $P4_2/mnm$ structure.
        The zero of energy is taken to be the Fermi energy.
  }
\end{figure}

During the relaxation calculation of the metallic tetragonal phase, we observed cases where the DFT relaxation was trapped in a local minimum with $a/c$ ratio above one.
All of these configurations had energies above the ground-state structure for that volume, but such local minima were seen for all of the halides.
One of the interesting structures was the $P4_2/mnm$ (No. 136) structure, as illustrated in Figures~\ref{fig:struct}(c) and (d).
The $P4_2/mnm$ (No. 136) structure is a subgroup of $P4/mbm$ (No. 127), the structure of \ch{KCuF3}, which is related by the phonon mode of the gold charge disproportionation.
Therefore we will call this structure as \ch{KCuF3} analog structure.
The \ch{KCuF3} analog structure was metastable at ambient pressure, and the relative energies above the ground state are 18.5, 14.5, and 23.0 meV/f.u. for X=Cl, Br, I, respectively.
The phase change upon pressurization is similar to that of the ground state $I4/mmm$ structure, and it undergoes tetragonal--cubic insulator--metal transition directly to the cubic $Pm\overline{3}m$ structure at 118, 141, and 168 ~\AA$^3$ for X=Cl, Br, and I, respectively.
Calculating the phonon modes, we found imaginary modes for $P4_2/mnm$ \ch{Cs2Au2Br6} and \ch{Cs2Au2I6} at the zone boundary, which would allow the structure to relax to the ground state $I4/mmm$ structure (Figure~\ref{figSI:band_phonon}).
On the other hand, the phonon eigenvalues for $P4_2/mnm$ \ch{Cs2Au2Cl6} are all real, suggesting it to be dynamically metastable.
Since complex interplay between pressure and phonons are reported experimentally,\cite{Trigo2012} we do not rule out the possibility that \ch{Cs2Au2Br6} and \ch{Cs2Au2I6} can also be kinetically stabilized under a certain strain environment, and therefore their results are also reported.

One of the known problems of double perovskites limiting their use for optoelectronic applications is their transport properties.
Even if the crystal structure appears favorable for three-dimensional transport, poor transport of electrons along all or some of the crystal axes is possible.
The concept of electronic dimensionality, advocated by Yan and co-workers, suggests that well-connected electronic density of the conduction or valence band edge states is a figure of merit along with band gap, for favorable transport properties and defect properties.\cite{XiaoZ2017_MH}
The electronic density of the (lower) conduction band for \ch{Cs2Au2Cl6} in $I4/mmm$ and $P4_2/mnm$ is shown in Figures~\ref{fig:band_edges}(a) and (c).
It is noticeable that some chlorine atoms lack significant electronic density in those phases.
These chlorines are the nearest neighbors of \ch{Au^{1+}}.
This is the consequence of \ch{Au^{1+}Cl6} octahedra having more orbitals occupied compared to the \ch{Au^{3+}Cl6} octahedra.
Thus, the lack of conduction band state on the chlorine adjacent to \ch{Au^{1+}} creates a disruption in the electronic dimensionality.
A schematic image of the electronic dimensionality for the $I4/mmm$ structure is shown in Figure~\ref{fig:band_edges}(b), which shows a two-dimensional network.
In contrast, the electronic dimensionality for the $P4_2/mnm$ structure is three-dimensional (Figure~\ref{fig:band_edges}(d)).
The \ch{Au^{1+}} ions (in all three dimensions) are distributed both in the $a$ and $c$ directions, such that an alternative path for electrons is possible.
The above discussion and features are also reflected in the band structure.
Between $\Gamma$-Z in $I4/mmm$ (Figure~\ref{fig:elec}), the dispersion is small, whereas in $P4_2/mnm$ (Figure~\ref{fig:elec_KCF}), the dispersion is large.

%
Although both the $I4/mmm$ and the $P4_2/mnm$ structures are built from the same \ch{Au^{1+}X6} and \ch{Au^{3+}X6} building blocks, their band structures show significant differences.
The stark difference is reflected in the band gap trends as the atomic number of the halide is increased, where it narrows in $I4/mmm$ (Figure~\ref{fig:elec}) but widens in $P4_2/mnm$ (Figure~\ref{fig:elec_KCF}).
The main conduction band is almost identical between the two structures and the valence band, despite having a different shape, results in a similar density of states.
Therefore, the difference is largely attributed to the lower conduction band.
Two main differences are present.
Firstly, in the ground-state $I4/mmm$ structure, the lower conduction band is almost identical across the halides, whereas in the \ch{KCuF3} analog $P4_2/mnm$ structure, they have a similar shape but the dispersion changes dramatically as the atomic number of the halide is increased.
The trend is explained by focusing on the distinctive ``eye glasses-like'' band structure (along $\Gamma$-Y-S-X-$\Gamma$) in $P4_2/mnm$.
Taking a look at the feature in the range $\Gamma$-Y-S, there are two sets of bands.
The horizontal bands are non-bonding and the dispersive bands are bonding.
The states in these isolated bands are constructed from the hybridization of Au and X orbitals (Figure~\ref{fig:elec_KCF}), and it can be suggested that the horizontal bands are more Au 5d-orbital-like, while the dispersive bands are more X p-orbital-like.
Therefore, the heavier the halide, the more dispersion is present.
Secondly, in $P4_2/mnm$, the dispersion in $\Gamma$-Y-S-X-$\Gamma$ and Z-T-R-U-Z have a similar shape, but the latter has a lower energy than the former.
This feature can be explained by the bonding network discussed in the previous paragraph.
In $I4/mmm$, the electronic structure is largely two-dimensional, so the coupling in the $c$ direction is weak, causing the dispersion to be similar in the $k_z=0$ and $k_z=\pi/2c$ planes, whereas in $P4_2/mnm$ the interaction in the $c$ direction is not small.
The two effects (dispersion and electronic dimensionality) collectively cause the bottom of the conduction band in \ch{Cs2Au2I6} to be lifted up compared to \ch{Cs2Au2Cl6}, causing the band gap to open.

The band structure of the \ch{KCuF3} analog $P4_2/mnm$ structure suggest that synthesizing structures with different octahedral orientations is a promising avenue to realize favorable electronic properties in double perovskites.
In spinful systems, the bonding network will directly influence the spin couplings and the orientation of the octahedra will enforce orbital ordering.
As seen in \ch{KCuF3},\cite{Paolasini2002,Scheie2021} this will also tune the spinon excitation patterns.
The localization of the conduction band has been raised as an obstruction to halide perovskite alloys for use in optoelectronic devices.\cite{XiaoZ2019}
The modification suggested in this study is a promising strategy to delocalize the band edge states.
We mainly looked into hydrostatic pressure as an external stimulus, but other methods such as adding defects, adding dopants, grain engineering, crystal morphology engineering, using different substrates, passivation and annealing in the synthesis process are additional practical methods to tune the octahedral distortion ordering.
The small energy difference between the different octahedral distortion orientations implies that many of these methods will be effective.

\section{Conclusion}
In summary, we investigated the electronic and structural property changes of \ch{Cs2Au2X6} (X=Cl, Br, I) under increasing pressure through DFT calculations.
The ground state tetragonal $I4/mmm$ turns into cubic $Pm\overline{3}m$ directly and is accompanied with insulator--metal transition.
We suggest that this should be the behavior in a clean, fully controlled experiment.
During this phase transition, in \ch{Cs2Au2Cl6}, the shortest bond length around \ch{Au^{3+}} increased with increasing pressure, which we suggested to be the consequence of small chlorine size.
We also observe that \ch{Cs2Au2Cl6} can be metastable in the \ch{KCuF3} analog $P4_2/mnm$ structure.
Although the $P4_2/mnm$ structure is composed of the same \ch{Au^{1+}X6} and \ch{Au^{3+}X6} building blocks, the behavior of the band gap upon increasing the atomic number was inverted relative to that of ground-state $I4/mmm$ structure.
We also found that the electronic dimensionality in $P4_2/mnm$ structure is increased to three dimensional from two dimensional in the ground state structure $I4/mmm$.
The experimental realization of \ch{Cs2Au2X6} in the $P4_2/mnm$ structure is yet to be seen, but our theoretical calculation supports the efficacy of the emerging strategy of using octahedral building block arrangement to design a new material.
Such an approach will greatly expand the material design space and allow the realization of unforeseen properties.

\begin{acknowledgement}
        K.M. and A.M.R. acknowledge support from the U.S. Department of Energy, Office of Science, Office of Basic Energy Sciences, under Award \#DE-FG02-07ER46431. Computational support was provided by the National Energy Research Scientific Computing Center (NERSC), a U.S. Department of Energy, Office of Science User Facility located at Lawrence Berkeley National Laboratory, operated under Contract No. DE-AC02-05CH11231.
        K.M. acknowledges the JSPS Overseas Research Fellowship.
\end{acknowledgement}

\begin{suppinfo}
        Details on the crystal structures, competing phases, HSE06 pressurization calculation, HSE06 band structures, bonding lengths, volume dependent pressure, and phonon dispersions, is provided in the Supporting Information.
\end{suppinfo}

\newpage
\bibliography{reference}

{\color{white}
%
\customlabel{tableSI:lattcon}{S1}
\customlabel{tableSI:lattcon_KCF}{S2}
\customlabel{tableSI:coord_KCF_Cl}{S3}
\customlabel{tableSI:coord_KCF_Br}{S4}
\customlabel{tableSI:coord_KCF_I}{S5}
\customlabel{figSI:struct_KCF_prim}{S1}
\customlabel{figSI:struct_prerelax}{S2}
\customlabel{figSI:ac_hse06}{S3}
\customlabel{figSI:band_HSE}{S4}
\customlabel{figSI:v_bond}{S5}
\customlabel{figSI:e_p}{S6}
\customlabel{figSI:band_phonon}{S7}
}

\end{document}